# The power-law reaction rate coefficient for the barrierless reactions

Yin Cangtao and Du Jiulin

*Department of Physics, School of Science, Tianjin University, Tianjin 300072, China*

**Abstract:** The power-law reaction rate coefficient for the barrierless reactions is studied if the reactions take place in systems with power-law distributions and a generalized rate formula for the barrierless reactions in Gorin model is derived. We show that, different from those for bimolecular and unimolcular reactions, due to barrierless the power-law rate coefficient for the barrierless reactions does not have the factor of power-law function and thus it is not very strongly dependent on the $\nu$-parameter. Four barrierless reactions are taken as application examples to calculate the new rate coefficients, which with larger fitting $\nu$-parameters can be exactly in agreement with measurements in the experimental studies.

**Keywords**：Reaction rate coefficient, power-law distribution, barrierless reaction, nonequilibrium system

## 1. Introduction

The phase space theory (PST) provides a useful and easily implemented reference theory for the barrierless reactions. The basic assumption in PST is that the interaction between two reacting fragments is isotropic and does not affect the internal fragment motions. Essentially, PST is one version of transition state theory (TST) that focuses on the energetics of the separated fragments at a completely loose transition state (i.e., the rotations of the fragments are completely unhindered). The simplest algorithms are based on locating the transition state at the centrifugal barriers for spherically symmetric $r^{-n}$ potentials. For example, for $n=4$ it is the model of Langevin-Gioumousis-Stevenson, and for $n=6$ it is the model of Gorin [1]. In Gorin model, the thermal rate coefficient is given [2] by

$$k_{Gorin}(T) = 2^{11/6} \Gamma(2/3) C^{1/3} \sqrt{\pi/\mu} (k_B T)^{1/6}, \tag{1}$$

where $\mu$ is the reduced mass of the collision partners, $C$ is a constant, $k_B$ is Boltzmann constant, and $T$ is temperature.

Like all other formations of TST, The formulae in PST are also based on the thermal equilibrium assumption [1-3]. However, in the reaction rate theory, what we are interested in is the evolution processes from one metastable state to another neighboring state of metastable equilibrium, and therefore the assumptions would be quite farfetched. A lot of theoretical works and experimental studies on physical, chemical, biological and technical processes taking place in complex systems have revealed that in many situations the statistical property of complex systems is not Boltzmann-Gibbs (BG) distributions, but often follows the power-law distributions (e.g., see [4] and the references therein). The power-law distributions in complex systems have been noted and studied prevalently in the processes such as single-molecule conformational dynamics [5,6], chemical reactions [7], gene expressions [8], cell reproductions [9], complex cellular networks [10], and small organic molecules [11] etc. In these processes, the reaction rate coefficients are often energy-dependent (and/or time-dependent [12-14]) with power-law forms [15, 16], which thus are beyond the scope of conventional reaction rate formulae with BG exponential forms. Under these situations, the rate formulae in the conventional reaction rate theories based on BG statistics have to be modified.

The statistical mechanical theory of power-law distributions has been developed. For example, the generalized Gibbsian theory for power-law distributions was presented to the systems away from equilibrium [17]. In stochastic dynamical theory on Brownian motion in a complex system, power-law distributions can be discovered by introducing generalized fluctuation-dissipation relations and solving Fokker-Planck equations [4,18]. It is especially worth mentioning that for recent years nonextensive statistical mechanics (NSM) based on Tsallis entropy has received great attention and very wide applications to a variety of interesting problems in physics, chemistry, astronomy, biology, engineering and technology [19]. NSM has also been the statistical base of the kappa-distributions observed in space plasmas [20-24]. In NSM, the power-law distribution can be derived using the extremization of Tsallis entropy. When one generalizes BG statistical mechanics to NSM, the usual exponential and logarithm can be replaced respectively by the $q$-exponential and the $q$-logarithm. Here we can write some of the distributions known in NSM. We can introduce the power-law energy $\nu$-distribution,

$$P(\varepsilon) \sim [1-(\nu-1)\varepsilon/(k_B T)]^{1/(\nu-1)}, \qquad (2)$$

if the energy $\varepsilon$ is small. Or we can write $P(\varepsilon) \sim \varepsilon^{-\alpha}$ if the energy $\varepsilon$ is large [4], where $\nu \neq 1$ is a parameter. This power-law $\nu$-distribution represents the statistical property of the systems being at a nonequilibrium stationary-state [24, 25]. Eq.(2) is reduced to a BG distribution if the $\nu$-parameter is set $\nu \to 1$, where the parameter $\nu \neq 1$ measures the distance away from thermal equilibrium.

Most recently, TST for nonequilibrium systems with power-law distributions was studied and the generalized reaction rate formulae of these cases were derived for one-dimensional and $n$-dimensional Hamiltonian systems [15]. And the power-law TST reaction rate coefficients for the elementary bimolecular reactions [26] and the unimolecular reactions [27], and the collision theory power-law rate coefficients [28] were also studied. These facts tell us that the energy distribution functions play a key role in the reaction rate theories if the reaction takes place in nonequilibrium complex systems. In addition, the mean first passage time [29] and the escape rate for power-law distributions in both overdamped systems [30] and low-to-intermediate damping [31] were studied, which showed new and important characteristics. As we can imagine, this is a complicated and exciting field in exploring the understanding of open nonequilibrium reaction rate theory.

The purpose of this work is to generalize the barrierless reaction rate formula to a nonequilibrium system with the power-law ν-distribution. In Sec.2, we study the power-law reaction rate coefficient of the barrierless reactions. In Sec.3, we make numerical analyses to show the dependence of the power-law reaction rate coefficient on the ν-parameter. As application examples of the new formula, in Sec.4 we calculate the power-law reaction rate coefficients of four barrierless reactions, and compare them with the rate values in the experiment studies, and then determine the ν-parameter. Finally, in Sec.5 we give the conclusion and discussion.

**2. The power-law rate coefficient for the barrierless reactions**

In this section, we will follow the standard line of textbooks to derive the power-law reaction rate formula. For the barrierless reactions, the relation between the state-selected rate constant $k_{ij}$ and the reaction cross section $\sigma_{ij}$ [2] is

$$k_{ij}(u_r) = u_r \sigma_{ij}(u_r), \tag{3}$$

where $u_r$ is the relative velocity between the two reactant particles. The average reaction cross section $\sigma_r$ is given by the average over all the reactants internal states,

$$\sigma_r = \sum_{i,j} w_i^A w_j^B \sigma_{ij}(u_r), \tag{4}$$

where $w_i^A$ and $w_j^B$ represent the weighting factors of $i$th and $j$th reactant internal states, respectively. The thermal rate constant for the process is defined by the average on $u_r \sigma_r$ over all $u_r$: $0 \sim \infty$,

$$k = \int_0^\infty \sigma_r u_r f(u_r) du_r, \tag{5}$$

where $f(u_r)$ is the relative velocity distribution function of the particles.

There are many methods to calculate the reaction cross $\sigma_r$. In PST, the effective potential method is an often used convenient method. The effective potential is given [2] by

$$V_{\text{eff}}(R) = V(R) + L^2/2\mu R^2, \tag{6}$$

where $R$ is a distance between the collision partners, $L$ is the orbital angular momentum, and $V(R)$ is the isotropic fragment-fragment interaction energy. For the neutral reactions without the barrier, the efficient potential in Gorin model [1, 2] is

$$V_{\text{eff}}(R) = -C/R^6 + L^2/2\mu R^2. \tag{7}$$

The first term in this equation is the ionization potential, and the second term is the centrifugal potential. Because of $L = \mu u_r b$, using the relative translational energy $E_r = \mu u_r^2/2$ we obtain

$$V_{\text{eff}} = -C/R^6 + E_r (b/R)^2, \tag{8}$$

Eq.(8) has a single maximum at a radius $R_* = (3C/E_r b^2)^{1/4}$ and the maximum is

$$V_{\text{eff},*} = 2E_r^{3/2} b^3 3^{-3/2} C^{-1/2}, \tag{9}$$

where $C$ is a constant determined by the nature of the molecules, and $b$ is an impact parameter defined as the distance between the closest two molecules in absence of inter-particle forces. If $E_r < V_{\text{eff},*}$, the centrifugal barrier cannot be penetrated when the tunneling is neglected, and no reaction occurs (in fact, the absence of a barrier also makes quantum tunneling effects usually not important for calculating the thermal rate constant [2] ). If $E_r = V_{\text{eff},*}$, one particle is captured into a circular orbit of radius $R_*$ around the other particle. If $E_r < V_{\text{eff},*}$, the two particles can move inside each other and the reaction probability is approximately equal to one. The critical impact parameter $b_*$ is obtained from $V_{\text{eff},*}$ by

$$b_* = 2^{-1/3} E_r^{-1/6} 3^{1/2} C^{1/6}, \tag{10}$$

and the reaction cross section is,

$$\sigma = \pi b_*^2 = 2^{-2/3} \pi E_r^{-1/3} 3 C^{1/3}. \tag{11}$$

Then the thermal rate coefficient is given by

$$k = \int_0^\infty du_r \sigma u_r f(u_r). \tag{12}$$

It is clear that this rate formula depends strongly on the relative velocity distribution function. Select the velocity distribution function depends on the statistical property of the systems under consideration. In the conventional theories, one used to presume that thermodynamic equilibrium has always been maintained in the systems. Under this assumption, the statistical distribution used to be naturally a BG distribution and therefore the relative velocity distribution has the exponential form [32],

$$f(u_r) = \left(\frac{\mu}{2\pi k_B T}\right)^{3/2} \exp\left(-\frac{\mu u_r^2}{2 k_B T}\right). \tag{13}$$

However, generally speaking, the chemical reaction is not always in a thermodynamic equilibrium with a BG distribution but usually in a nonequilibrium state with non-BG distributions. In the reaction rate theory, what we are interested in is the processes of the evolution from one metastable state to another neighboring state, thus the thermodynamic equilibrium assumption is quite farfetched. In particular, in some complex systems, the system far away from equilibrium does not relax to a thermodynamic equilibrium with a BG distribution, but often asymptotically approaches a stationary nonequilibrium with a power-law distribution. In this case, the relative velocity distribution function, Eq.(13), is not true and consequently the reaction rate coefficient Eq.(1) should be modified.

In NSM, the power-law distribution (2) can be derived using the extremization of Tsallis entropy. When BG statistics is generalized to nonextensive statistics, the usual used exponential and logarithm can be replaced respectively by the corresponding $q$-exponential and $q$-logarithm [19]. Here the $\nu$-exponential [33, 34] can be defined as

$$\exp_\nu x = [1 + (\nu - 1)x]^{1/(\nu - 1)}, \quad (\exp_{\nu=1} x = e^x), \tag{14}$$

if $1 + (\nu - 1)x > 0$ and as $\exp_\nu x = 0$ otherwise. And the inverse function, the $\nu$-logarithm can be defined as

$$\ln_\nu x = \frac{x^{\nu-1}-1}{\nu-1}, \quad (x>0, \ln_{\nu=1} = \ln x). \tag{15}$$

In this framework, using the energy distribution (2) the relative velocity distribution (13) can be generalized to the power-law $\nu$-distribution,

$$f_\nu(u_r) = Z_\nu \left[1-(\nu-1)\frac{\mu u_r^2}{2k_B T}\right]^{1/(\nu-1)}, \tag{16}$$

where $Z_\nu$ is the normalization constant [35],

$$Z_\nu = \left(\frac{m}{2\pi k_B T}\right)^{\frac{3}{2}} \begin{cases} (\nu-1)^{\frac{5}{2}} \Gamma\left(\frac{1}{\nu-1}+\frac{5}{2}\right) \Big/ \Gamma\left(\frac{1}{\nu-1}\right), & \text{if } \nu>1, \\ (1-\nu)^{\frac{3}{2}} \Gamma\left(\frac{1}{1-\nu}\right) \Big/ \Gamma\left(\frac{1}{1-\nu}-\frac{3}{2}\right), & \text{if } 1/3<\nu<1. \end{cases} \tag{17}$$

In the limit $\nu \to 1$, (16) is reduced to (13), the form in conventional BG thermodynamic statistics. Here, it would be helpful to introduce the physical meaning of the power-law parameter $\nu \neq 1$. In 2004, an equation of the parameter $\nu \neq 1$ was found both in the self-gravitating and plasma systems with the long-range interactions and hence a clear physical explanation for $\nu \neq 1$ was presented [24, 25]. The equation can be written as $k_B \nabla T(r) = -(\nu-1)m\nabla \varphi_g(r)$ for the self-gravitating system [25] and $k_B \nabla T(r) = (\nu-1)e\nabla \varphi_C(r)$ for the plasma system [24], where $T(r)$ is space-dependent temperature, $m$ is mass of particle, $e$ is charge of electron, $\varphi_g(r)$ is a gravitational potential function, and $\varphi_C(r)$ is a Coulombian potential function. The equation shows that the $\nu$-parameter is $\nu \neq 1$ if and only if $\nabla T(r) \neq 0$ and hereby the power-law distribution represents the nature of an interacting many-body system being at a nonequilibrium stationary-state. For a chemical reaction system, the equation of the $\nu$-parameter should be similar to that for the self-gravitating system (although one needs to study the precise expression), and $\varphi_g(r)$ should be construed as an intermolecular interaction potential function.

Further, substituting Eq. (16) into Eq. (12), one can find that the reaction rate coefficient in the Gorin model becomes $\nu$–dependent,

$$k_\nu = 2^{-\frac{2}{3}} 12\pi^2 C^{\frac{1}{3}} Z_\nu \int_0^\infty du_r E_r^{-\frac{1}{3}} u_r^3 \left[1-(\nu-1)\frac{E_r}{k_B T}\right]^{1/(\nu-1)}. \tag{18}$$

After completing the integration in Eq.(18), we can derive the reaction rate coefficient for the barrierless reaction in the system with power-law velocity $\nu$-distribution (i.e., the power-law rate coefficient),

$$k_\nu = 2^{\frac{11}{6}} \Gamma\left(\frac{2}{3}\right) \sqrt{\frac{\pi}{\mu}} C^{\frac{1}{3}} (k_B T)^{\frac{1}{6}} K_\nu, \tag{19}$$

with the $\nu$-dependent factor,

$$K_\nu = \begin{cases} (\nu-1)^{-\frac{1}{6}} \Gamma\left(\frac{1}{\nu-1}+\frac{5}{2}\right) \Big/ \Gamma\left(\frac{1}{\nu-1}+\frac{8}{3}\right), & \text{if } \nu>1, \\ (1-\nu)^{-\frac{1}{6}} \Gamma\left(\frac{1}{1-\nu}+\frac{5}{3}\right) \Big/ \Gamma\left(\frac{1}{1-\nu}-\frac{3}{2}\right), & \text{if } \frac{2}{5}<\nu<1. \end{cases} \qquad (20)$$

As compared with the old formula, the new rate coefficient Eq.(19) has a $\nu$–dependent factor $K_\nu$, but does not have the factor of power-law $\nu$-distribution due to barrierless. As expected, in the limit $\nu \to 1$ it is reduced to the standard reaction rate formula for the reactions in systems with a BG distribution [1, 2],

$$k_1 = 2^{\frac{11}{6}} \Gamma\left(\frac{2}{3}\right) \sqrt{\frac{\pi}{\mu}} C^{\frac{1}{3}} (k_B T)^{\frac{1}{6}}. \qquad (21)$$

**3. Numerical analyses of the power-law barrierless reaction rate coefficient**

In order to illustrate the characteristics of the power-law rate coefficient $k_\nu$ in Eq.(19) and show the dependence of $k_\nu$ on the $\nu$-parameter and the temperature, we make numerical analyses.

Fig.1 showed the dependence of the rate coefficient $k_\nu$ on the parameter $\nu$. The $k_\nu/k_1$-axis was plotted on a logarithmic scale. The range of the $\nu$-axis was chosen from 0.5 to 3.0, a typical range of the values in nonextensive statistics. It was shown that the rate coefficient $k_\nu$ decreases as the parameter $\nu$ increases, which implied that a deviation from a BG distribution and thus from thermal equilibrium would result in a significant variation in the reaction rate. However, we find that such a $\nu$–dependent variation in the reaction rate for the case $\nu>1$ and the case $\nu<1$ is different. If $\nu>1$, the reaction rate will decrease as the $\nu$-parameter deviates from one (i.e., the system deviates from thermal equilibrium). But if $\nu<1$, the reaction rate will increase as the $\nu$-parameter deviates from one (i.e., the system deviates from thermal equilibrium). This characteristic comes from the different physical states represented by the system for the two cases $\nu>1$ and $\nu<1$, respectively. For the deep understanding of this characteristic, we may need to find the exact expression of the $\nu$-parameter of a chemical reaction system.

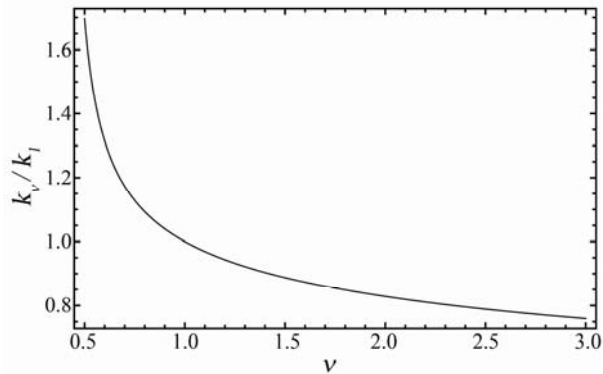

Fig. 1. Dependence of $k_\nu/k_1$ on the parameter $\nu$

Fig.2 illustrated the dependence of the rate coefficient $k_\nu$ on the temperature $T$ for three different $\nu$-parameters. The range of $T$-axis was chosen as 100 ~ 1000K, the typical temperature range in chemical reactions. The line of $\nu = 1$ is corresponding to the conventional reaction rate coefficient $k_1$. It was shown that the power-law rate coefficient increases as the temperature increases and there are differences, but not very significant, for different $\nu$-parameters about $\nu=1$.

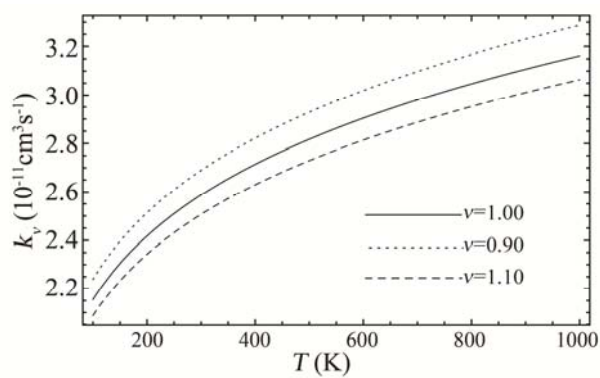

Fig. 2. Dependence of $k_\nu$ on temperature $T$ for three values of $\nu$

Nevertheless, different from the power-law rate coefficients for the bimolecular reactions [26], the unimolecular reactions [27] and the collision theory [28], because there are not the barriers the power-law rate coefficient Eq.(19) for the barrierless reactions does not have the factor of power-law $\nu$-distribution and thus it is not very strongly dependent on the $\nu$-parameter, as compared with those for the bimolecular and unimolecular reactions, and the collision theory.

## 4. Application of the new rate coefficient to barrierless reactions

In order to illustrate the application of the new reaction rate formula Eq.(19) to the barrierless reactions which occur in a nonequilibrium system with the power-law $\nu$-distribution, we take four barrierless reactions as below in table 1 to calculate the power-law reaction rate coefficients and compare them with the measurements in the experimental studies.

Table 1. The rate coefficients for four barrierless reactions

| Reactions | $k_1(\text{cm}^3\text{s}^{-1})$ | $k_{\text{exp}}(\text{cm}^3\text{s}^{-1})$ | $\delta$ | $k_\nu(\text{cm}^3\text{s}^{-1})$ | $\nu$ |
|---|---|---|---|---|---|
| ·CH$_3$ + ·CH$_3$ | 1.05×10$^{-10}$ | 4.03×10$^{-11}$ | 160% | 4.03×10$^{-11}$ | 147 |
| ·CH$_2$Cl$_3$ + ·CH$_2$Cl$_3$ | 8.32×10$^{-11}$ | 3.43×10$^{-11}$ | 142% | 3.43×10$^{-11}$ | 97 |
| ·CCl$_3$ + Cl | 8.28×10$^{-11}$ | 7.29×10$^{-11}$ | 14% | 7.29×10$^{-11}$ | 1.55 |
| C$_2$H$_4$ + Cl | 3.31×10$^{-10}$ | 2.55×10$^{-10}$ | 30% | 2.55×10$^{-10}$ | 2.82 |

In Table 1, we listed the experimental values and the theoretical values of the rate coefficients for these four reactions. $k_1$ is the conventional rate coefficient calculated by using Eq.(21), $k_\nu$ is the power-law rate coefficient calculated by using Eq.(19), and $k_{\text{exp}}$ is the measured rate coefficient in the experimental studies taken from the NIST chemical kinetics database at http://kinetics.nist.gov/kinetics,. The quantity $\delta$ denotes the relative error of $k_1$ to $k_{exp}$, i.e. $\delta=|k_1 - k_{exp}| / k_{exp}$, and $\nu$ is the fitting power-law parameter. All the data were obtained at the temperature 400K.

We find that there are very significant relative errors of $k_1$ to $k_{exp}$, and the values of $k_\nu$ with different and larger $\nu$-parameter can be exactly in agreement with all the experimental studies. New rate coefficient $k_\nu$ with larger fitting $\nu$-parameters shows that due to barrierless the power-law rate coefficient for the barrierless reactions is not very sensitive to the $\nu$-parameter. In other words, the reaction rate is not strongly dependent on the form of the energy distribution function, and thus the reason for the significant relative errors of $k_1$ to $k_{exp}$ might need to search other explanations besides the nonextensive effect.

## 5. Conclusion

The reaction rate theory for the reactions in systems with power-law distributions is beyond the scope of the conventional theories with a BG distribution and therefore the reaction rate formulae need to be modified for the systems with power-law distributions. In this work, we have studied the power-law reaction rate coefficient for the barrierless reactions in Gorin model and derived the generalized reaction rate formula Eq.(19). We show that, different from the power-law rate coefficients for the bimolecular and unimolecular reactions, because there are not the barriers the power-law rate coefficient (19) for the barrierless reactions does not have the factor of power-law $\nu$-distribution and thus it is not very strongly dependent on the $\nu$-parameter, as compared with those for the bimolecular and unimolecular reactions.

We have made numerical analyses to illustrate the dependence of the power-law reaction rate coefficient $k_\nu$ on the power-law $\nu$-parameter.

We have also taken four barrierless reactions as the application examples of Eq.(19) to calculate the power-law rate coefficients. We showed that new rate coefficients with larger fitting $\nu$-parameters can be exactly in agreement with the measurements in the experimental studies. Again, due to barrierless the power-law rate coefficient $k_\nu$ for the barrierless reactions is not very sensitive to the $\nu$-parameter and thus the reaction rate is not strongly dependent on the form of the energy distribution function. And the reason for significant relative errors of the conventional rate coefficient $k_1$ to the measured rate coefficient $k_{exp}$ might need to search for other explanations besides the nonextensive effect.

**Acknowledgment**

This work is supported by the National Natural Science Foundation of China under Grant No. 11175128 and also by the Higher School Specialized Research Fund for Doctoral Program under Grant No. 20110032110058.